\documentclass[preprints,article,accept,pdftex,moreauthors]{Definitions/mdpi} 

\usepackage{caption}
\usepackage{subcaption}
\pdfoutput=1

\firstpage{1} 
\pubvolume{1}
\issuenum{1}
\articlenumber{0}
\pubyear{2022}
\copyrightyear{2022}
\externaleditor{Biagio Lucini, Chik Him Wong, Maarten Golterman}
\datereceived{} 
\dateaccepted{} 
\datepublished{} 
\hreflink{https://doi.org/} 

\usepackage{textcomp}

\newcommand{\I}{\ensuremath{\mathrm{i}}}

\newcommand{\mg}{m_{\tilde{g}}}

\Title{Exploring gauge theories with adjoint matter on the lattice}

\TitleCitation{Title}

\Author{Georg Bergner $^{1,2}$\orcidA{}, Gernot Münster$^{2}$\orcidB{}, and Stefano Piemonte $^{3}$\orcidC{}}

\AuthorNames{Georg Bergner, Gernot Münster, and Stefano Piemonte}

\AuthorCitation{Bergner, G.; Münster, G.; Piemonte, S.}

\address{%
$^{1}$ \quad University of Jena, Institute for Theoretical Physics, Max-Wien-Platz 1, D-07743 Jena, Germany; georg.bergner@uni-jena.de\\
$^{2}$ \quad University of M\"unster, Institute for Theoretical Physics, 
Wilhelm-Klemm-Str.~9, D-48149 M\"unster, Germany\\
$^{3}$ \quad University of Regensburg, Institute for Theoretical Physics, 
Universit\"atsstr.~31, D-93040 Regensburg, Germany; stefano.piemonte@ur.de}

\abstract{We review our efforts in investigating gauge theories with fermions in the adjoint representation of the gauge group by means of numerical simulations. These theories have applications in possible extensions of the Standard Model of particle physics, being a core part of supersymmetric gauge theories. They also play an important role in uncovering fundamental properties of strongly interacting theories due to distinct features, such as a substantially different phase diagram.}

\begin{document}
\section{Introduction}
Our understanding and view of strong interactions of elementary particles is primarily shaped by the knowledge we have gained about quantum chromodynamics (QCD). Experimental evidence, effective theories, and numerical lattice investigations have provided a lot of information about QCD despite the fact that the strongly coupled regime still withstands an analytic solution. QCD is an SU(3) gauge theory with fermions in the fundamental representation, but in general a field content with fermionic and scalar fields in various representations of the gauge group can be coupled to gauge fields. Considering a more general matter content provides new perspectives on strong interactions, and new insights are obtained from larger symmetries that can be implemented in such a generalized theory. Such kind of theories can also provide strongly interacting sectors beyond the Standard Model of particle physics.

In this article, we focus on gauge theories with fermions in the adjoint representations and different additional matter content. Fermions in the adjoint representation are described by spinor fields $\psi^{a}(x)$, where the index $a$ runs from 1 to $N_c^2 - 1$ in the case of gauge group SU($N_c$). They can be collected in a matrix valued field $\psi(x) = \psi^{a}(x) T^{a}$, where $T^{a}$ are the generators of the gauge group. As we will explain below, choosing the adjoint representation leads to theories which share many similarities with QCD, but have, at the same time, quite distinct features and properties. This makes these theories interesting candidates for theoretical considerations and numerical investigations.

One of the distinct features of gauge theories with adjoint matter is the close relation to supersymmetric gauge theories. As the gauge bosons transform in the adjoint representation, their superpartners need to be fermions in the same representation. Any supersymmetric gauge theory therefore contains fermions in the adjoint representation. The simplest model is pure $\mathcal{N}=1$ supersymmetric Yang-Mills theory (SYM) with a single Majorana fermion in the adjoint representation. In supersymmetric QCD supermultiplets are added containing fermionic quark and scalar squark fields in the fundamental representation. In $\mathcal{N}=2$ and $\mathcal{N}=4$ supersymmetric Yang-Mills theory the matter content is, like the gauge field, in the adjoint representation. Supersymmetry has provided quite unique insights into non-perturbative properties using analytic methods. Conjectures have been made that postulate more general relevance of these findings, but non-perturbative numerical calculations are the only way to verify it.

Supersymmetric theories are candidates for an extension of the Standard Model. Since supersymmetry is not realized in physics at low energies, it must be broken, and the understanding of any non-perturbative breaking mechanism requires investigations on the lattice. Theories with adjoint matter are in a much more general sense relevant for possible extensions of the Standard Model. Prominent examples are composite Higgs models, but these theories have also been considered as possible candidates for quite generic dark matter models.

\section{Supersymmetric Yang-Mills theory}
The simplest supersymmetric model with gauge interactions and with the minimal adjoint fermionic matter content is $\mathcal{N}=1$ SYM. Since the adjoint representation is real, the minimal matter content, the gluino, is a Majorana fermion which matches on-shell the degrees of freedom of the bosonic gluon. The gluino field $\lambda(x) = \lambda^{a}(x) T^{a}$ obeys the Majorana condition $\bar{\lambda}(x) = \lambda^{T}(x) C$ with the charge conjugation matrix $C$, thus gluinos are their own antiparticles.

In Minkowski space, the on-shell Lagrangian of the theory is
\begin{equation}
\mathcal{L}_{\text{SYM}} = -\frac{1}{4} F^a_{\mu\nu} F^{a,\mu\nu} 
+ \frac{\I}{2} \bar{\lambda}^a \gamma^\mu \left( \mathcal{D}_\mu \lambda \right)^a 
- \frac{\mg}{2} \bar{\lambda}^a \lambda^a .
\end{equation}
where $F^a_{\mu\nu}$ is the non-Abelian field strength tensor, and $\mathcal{D}_\mu$ is the covariant derivative in the adjoint representation of the gauge group, given by $(\mathcal{D}_{\mu} \lambda)^{a} = \partial_{\mu} \lambda^{a} + g\,f_{abc} A^{b}_{\mu} \lambda^{c}$. The Lagrangian also includes a gluino mass term with mass $\mg$, which is necessary in view of the numerical simulations. For $\mg \neq 0$ this term breaks supersymmetry softly, which means that it does not affect the renormalization properties of the theory and that the spectrum of the theory depends on the gluino mass in a continuous way.

A large number of analytic predictions have been derived for this theory, like the exact beta function \cite{Novikov:1983uc}, the gluino condensate $\bar{\lambda} \lambda$ \cite{Hollowood:1999qn}, and low energy effective actions \cite{Veneziano:1982ah,Farrar:1997fn,Farrar:1998rm}. It exemplifies how supersymmetry leads to additional insights not possible for other theories.

The particle spectrum of a supersymmetric theory should be composed out of supersymmetry multiplets. The simplest multiplet in four dimensions is a chiral multiplet consisting of a scalar, a pseudoscalar, and a fermionic spin-\textonehalf~state. The complete eigenvalue spectrum of the Hamiltonian should have degenerate bosonic and fermionic states. This does not imply cancellations in the thermal ensemble since the fermions and bosons obey different statistics, but there are still several interesting aspects of finite temperature SYM, which will be discussed below. The cancellation between fermionic and bosonic states becomes relevant in a twisted partition function, which is the Witten index \cite{Witten:1982df}.

A prominent generic feature of supersymmetry is its close connection to space-time symmetries, implying that certain correlation functions are constant in supersymmetric theories and long distance physics is determined by short distance behavior \cite{Amati:1988ft}. The Hamilton operator representing time translations is also connected to the supersymmetry algebra. This leads to the fact that the Witten index, a twisted partition function summing over the difference between the contributions of bosonic and fermionic energy levels, becomes invariant under changes of parameters such as the space-time volume or the gauge coupling. In the path integral formulation the Witten index is represented by a compactified theory with periodic boundary conditions in the Euclidean time direction for fermions. The Witten index is determined by the bosonic and fermionic ground states, which implies that ground state physics is still dominant at small compactifications of the theory in the twisted partition function. These brief considerations exemplify how semiclassical physics survives from short distances to long distances.

Our collaboration has devoted a long term project to the simulations of $\mathcal{N}=1$ SYM. We have successfully studied the most interesting non-perturbative properties of this theory, which are the spectrum of particle bound states and the phase transitions. Only numerical simulations can provide results for these properties beyond perturbation theory and semiclassical studies. 

One particular advantage of $\mathcal{N}=1$ SYM is the rather simple tuning towards the supersymmetric continuum limit. It has been shown that only a single parameter, the bare gluino mass, has to be tuned \cite{Curci:1986sm,Suzuki:2012pc}. Up to lattice artefacts that vanish in the continuum limit, the same tuning can be applied for chiral symmetry and supersymmetry, as explained in a seminal paper by Curci and Veneziano \cite{Curci:1986sm}. Essentially, the fine-tuning problem of supersymmetry can be traced back to the tuning of chiral symmetry, as is commonly done in simulations of QCD. In fact, like in QCD, the bare mass has to be adjusted according to the signals of chiral symmetry if Wilson fermions are used, whereas no tuning is required for Ginsparg-Wilson fermions \cite{Ginsparg:1981bj}. Simulations with Ginsparg-Wilson fermions require intense numerical computations. Therefore they are not practicable for investigations of the spectrum of bound states. Ginsparg-Wilson fermions provide, however, a clearer definition of chiral symmetry breaking and are hence profitable for investigations of the gluino condensate.

\subsection{The particle spectrum of supersymmetric Yang-Mills theory}
The first and most important property characterizing $\mathcal{N}=1$ SYM is the spectrum of bound states, which has been a subject of our long term investigations, first for the gauge group SU(2) and later also for SU(3). We have used the Curci-Veneziano approach and tuned the bare gluino mass according to the signals of restored chiral symmetry in simulations with improved and unimproved Wilson fermions. This approach allowed us to generate large enough ensembles of gauge configurations to measure the correlation functions of the bound states. 

The spectrum of the theory contains glueball states, like in pure Yang-Mills theory. In addition there are meson-like gluino-balls and mixed gluino-glueballs, which can be formed in a theory with adjoint fermions. The scalar and pseudoscalar boson fields in the lightest chiral multiplet have first been conjectured to be meson-like gluino-balls \cite{Veneziano:1982ah}. Later on glueball states have been considered \cite{Farrar:1997fn}, followed by some further reasoning about the possible mixing of these states, see e.~g.~\cite{Feo:2004mr}. The fermionic field is believed to be in the form of a gluino-glue operator.

The numerical signal for all of these states is, unfortunately, rather noisy. This can be understood from the QCD counterparts, which are glueball states and flavor singlet mesons. This is the basic reason why the cost intensive Ginsparg-Wilson fermions are so far not optimal for this investigation. 

In QCD, precise signals for chiral symmetry are obtained from the pion mass or the partially conserved axial current relations. In SYM, chiral symmetry is broken to a discrete subgroup only (see Section~\ref{sec:gluinocond}), and the theory does not contain pseudo-Goldstone bosons. It is, however, possible to define an unphysical pion in partially quenched chiral perturbation theory \cite{Munster:2014cja}. Its correlation function is given by the fermion-connected part of the correlation function of the pseudoscalar gluino-ball, and can be measured numerically. From its decay with distance, the so-called adjoint pion mass is obtained, whose vanishing represents a signal of chiral symmetry. The adjoint pion mass is an easily measurable quantity that can be employed for tuning to the chiral limit. This approach is not ideal since it relies on an unphysical particle. As an alternative, the supersymmetric Ward identities can be used to determine the chiral limit. We checked that the tuning by means of the adjoint pion mass is consistent with an approach using the supersymmetric Ward identities \cite{Ali:2018fbq}. Yet another signature for chiral symmetry and its breaking is the histogram of the chiral condensate, which will be discussed in more detail below. For practical purposes, the difference between these tuning signals does not play a major role. We confirmed that they all lead (within the uncertainties) to a consistent picture, which means that possible differences disappear in the continuum limit.

The static quark-antiquark potential for fundamental sources shows a clear linear rise at large distances and, as expected, the running of the gauge coupling is QCD-like from asymptotic freedom to confinement. Therefore the parameters $r_0$, $w_0$ \cite{Sommer:1993ce,Borsanyi:2012zs} and other observables can be used in the same way as in QCD to determine a scale. In most cases, we used the $w_0$ scale, which can be determined rather accurately, but we have also determined $r_0$ and the static potential.

Pioneering work about the methods and algorithms was performed in \cite{Montvay:1995ea,Donini:1997hh}.
In the early data from simulations of SU(2) SYM, a multiplet formation could not be confirmed \cite{Campos:1999du}, which has lead to puzzles outside the lattice community \cite{Cerdeno:2003us}. Indeed, a large gap in the spectrum has been observed even in later simulations with improved algorithms \cite{Demmouche:2010sf}. One important lesson to be learned from these first studies is that the parameters and scales of theories might be different from expectations based on the present long experience with numerical simulations of QCD. The fact that lattice investigations of Yang-Mills theory and QCD can count on a very long history with a large number of contributions from different collaborations is often not well enough recognized.

Finally we were able to perform a larger number of investigations of SU(2) SYM, checking finite size effects \cite{Bergner:2012rv} and lattice artefacts. The complete continuum extrapolation shows that the gap between fermionic and bosonic states closes and one obtains a chiral multiplet in the complete continuum limit \cite{Bergner:2015adz,Ali:2019gzj}.

\begin{figure}
\begin{center}
\input{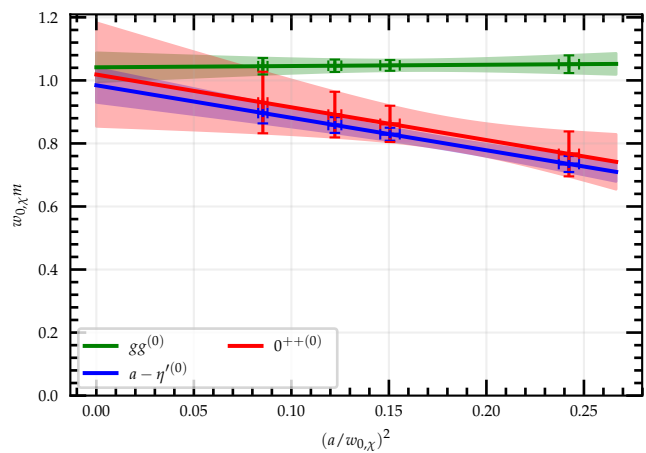}
\end{center}
\caption{Bound state masses of SU(3) SYM in a simultaneous extrapolation to the chiral and continuum limits ($a \to 0$). The lightest bound states are expected to form a chiral multiplet consisting of a scalar $0^{++}$, a pseudoscalar $a-\eta'$, and a fermionic gluino-glue ($gg$) state. The upper index $(0)$ indicates that it is the lightest state in each channel. Further details about the extrapolations and data are presented in \cite{Ali:2019agk}.} 
	\label{fig:continuumextr}
\end{figure}

We have continued our studies with the gauge group SU(3), which before had been considered in a first test \cite{Feo:1999hw}, but later on discarded due to the additional computational cost. With the experience from the successful continuum extrapolations of SU(2) SYM, we were able to select a good parameter region and lattice action to simulate SU(3) SYM efficiently. The final results in \cite{Ali:2018dnd,Ali:2019agk} confirmed again a multiplet formation, see Figure \ref{fig:continuumextr}. The consistency of the chiral tuning with supersymmetric Ward identities has also been confirmed \cite{Ali:2020mvj}.

Further interesting results for SU(3) SYM using a different lattice action can be found in \cite{Steinhauser:2020zth}. The large $N_c$ limit of SU($N_c$) SYM has been simulated in \cite{Butti:2022sgy}. Another approach for simulations of SYM using orientifold planar equivalence in the large $N_c$ limit has been proposed in \cite{Ziegler:2021nbl}. Simulations with domain wall and overlap fermions will be reported below in the discussion of chiral symmetry breaking. All of these works are highly appreciated since it is important to confirm findings not only by one collaboration using a single approach with only a certain lattice action.

The simulations require solutions of some additional technical problems, which we will only briefly mention here. 
Some problems are known from QCD simulations, like the glueball mass determination, the mixing of meson and glueball states, and the calculation of fermionic disconnected contributions. Different from QCD is the presence of Majorana fermions, which must be considered in the fermion measure of the path integral. They lead to a Pfaffian $\textrm{Pf}(M)$ instead of a determinant $\textrm{det}(M) = (\textrm{Pf}(M))^2$ of the Dirac operator. Up to a sign factor, the Pfaffian can be taken into account by a rational or polynomial hybrid Monte-Carlo algorithm. The sign needs to be added by reweighting. The sign of the Pfaffian, which is the reweighting factor, is obtained from the eigenvalues of the Dirac operator \cite{Bergner:2011zp}.

The sign fluctuations with Wilson fermions increase towards the chiral limit, but are reduced towards the continuum limit. In the investigations of the particle spectrum it is therefore always possible to approach the continuum limit in such a way that the sign problem does not become relevant. This is, however, not always the case in investigations of the phase transition discussed below.

\subsection{Zero temperature phases and the gluino condensate}
\label{sec:gluinocond}
SU($N_c$) $\mathcal{N}=1$ SYM is expected to have $N_c$ different vacuum states, each of them labeled by different possible values of the gluino condensate. The number of vacuum states can be considered to be the result of the breaking of chiral symmetry. The axial anomaly for fermions in the adjoint representation breaks the U(1)$_A$ symmetry (R-symmetry in SYM) to a discrete $Z_{2N_c}$ subgroup \cite{Curci:1986sm}, which rotates $N_c$ values of the fermion condensate into each other. Finally, it is this remaining discrete chiral symmetry that is spontaneously broken by the formation of the fermion condensate, and therefore, unlike QCD, we do not expect to observe Goldstone bosons associated with this symmetry breaking. The vacuum of $\mathcal{N}=1$ SYM is therefore expected to represent the coexistence of regions with different values of the chiral condensate, and with domain walls separating them. The spontaneous breaking of chiral symmetry is a pure non-perturbative phenomenon, and lattice simulations are ideal to investigate whether it occurs. However, the Nielsen-Ninomiya theorem limits the studies of chiral symmetry on the lattice.

Wilson fermions break chiral symmetry explicitly, and we are forced to renormalize the fermion condensate additively and multiplicatively. The additive renormalization prevents us from being able to distinguish the phases where the condensate is zero from the phases where chiral symmetry is broken in a simple manner. We have explored three different approaches to determine the value despite these difficulties. First, remember that the fermion mass plays a similar role as the external magnetic field in the Ising model, forcing the fields to be aligned in one direction. There is a first order transition crossing the chiral limit going from positive to negative values of the renormalized fermion mass, and the vacuum expectation value of the condensate should jump. The histogram should indicate a two peak structure corresponding to the two different vacua at this point for the gauge group SU(2), and the magnitude of the jump would provide the value of the bare chiral condensate free of additive renormalization. Early investigations of this approach can be found in \cite{Kirchner:1998mp} for SU(2) and in \cite{Feo:1999hw} for SU(3) SYM. In the case of SU(3), we also expect to observe a phase where the pseudoscalar condensate is non-zero, building effectively a three-peak state in the complex plane with the scalar and pseudoscalar fermion condensates in the real and imaginary axis, respectively. So far we have been unable to observe this peak structure in the imaginary axis, having seen a signal only in the scalar condensate \cite{Ali:2018dnd}. The positive result is nevertheless that the value of the fermion mass where we observe a double-peak structure is consistent with the signals provided in the partially quenched theory mentioned above, meaning that the two methods are consistent. 

The study of the coexistence of many phases has significant limitations. The signal is not very precise, it is difficult to extend the study to large volumes, and the simulations become unstable at the chiral point. A different approach to avoid the additive renormalization of the fermion condensate with Wilson fermions is provided by the gradient flow. The gradient flow is a smoothing acting on fermion and gauge fields, providing a regularization in addition to the lattice spacing \cite{Luscher:2010iy}. In this way, different lattice actions and different regularizations can be compared on the same footing, and in particular the chiral condensate is free from additive renormalization. We have used the gradient flow to extrapolate a meaningful non-zero value of the gluino condensate in the chiral limit \cite{Bergner:2019dim}, thereby directly confirming chiral symmetry breaking.

The gluino condensate can be better investigated on the lattice if a properly formulated (modified) chiral symmetry is preserved even for non-zero values of the lattice spacing. Following Ginsparg and Wilson \cite{Ginsparg:1981bj}, a modified chiral symmetry means that naive chiral rotations in the continuum are modified by additional terms proportional to the lattice spacing and vanishing in the continuum limit. This can be implemented by means of Ginsparg-Wilson fermions. The corresponding Dirac operators allow to properly link their zero eigenmodes to the topology of lattice gauge-field configurations, an important feature required to understand the role of semiclassical objects in gluino condensation. Ginsparg-Wilson fermions are however computationally quite demanding. The value of the condensate has been obtained from studies with Domain-Wall fermions \cite{Fleming:2000fa,Endres:2009yp,Giedt:2008xm}, and first preliminary results with overlap fermions have appeared in \cite{Kim:2011fw}. We have implemented a polynomial approximation of the overlap operator. Interestingly, our final result for the gluino condensate is in rough agreement with the results from gradient flow and Domain wall fermions, a precise matching can be possible only when the multiplicative renormalization is fixed by choosing a common renormalization scheme. 


\subsection{Phase transitions in supersymmetric Yang-Mills theory}
As we have seen, SYM shares some similarities with QCD. The deconfinement phase transition shows, however, very significant differences between the two theories. In QCD there is an explicit breaking of center symmetry by the fermions in the fundamental representation. By contrast, adjoint fermion fields are consistent with center symmetry, and in the chiral limit also chiral symmetry is restored. Therefore deconfinement is a phase transition and not a crossover for any value of the fermion mass, and dynamical chiral symmetry breaking can be observed once the renormalized fermion mass tends to zero. In contrast to QCD there is hence a meaningful question concerning the relationship between chiral transition (critical temperature $T_\chi$) and deconfinement transition (critical temperature $T_d$). The 't Hooft anomaly matching conditions provide constraints on the ordering of the transitions ($T_\chi>T_d$).

The interplay between the chiral and the deconfinement transition has been a motivation for several numerical studies. None of these studies has considered SYM. Instead they were focused on a larger number of fermion fields in the adjoint representation. In the earliest papers a huge difference between chiral and deconfinement transition has been observed with $T_\chi$ of the order of $170 T_d$ \cite{Kogut:1985xa}. In later studies this difference has been reduced to around $(7-8) T_d$, which is still a large difference \cite{Karsch:1998qj,Engels:2005te}.

$\mathcal{N}=1$ supersymmetric Yang-Mills theory provides a cleaner setup for an investigation of the two phase transitions. As will be explained in later sections, it is far enough away from the conformal window to avoid large scale finite size effects and bulk phases. In addition, supersymmetry has led to additional insights into the expected phase structure and to conjectures about the phase transitions. Since the main additional difficulties of simulating this theory have now been solved, it is therefore the ideal starting point for studies of the relation between the chiral and deconfinement transitions.

We have done several investigations to determine the chiral and the deconfinement transitions at non-zero temperatures \cite{Bergner:2014saa,Bergner:2019dim}. In an approach with a fixed number $N_t$ of lattice points in temporal direction, we have varied the temperature by changing the lattice spacing. In addition we have done a fixed scale analysis, in which the bare parameters are kept fixed and the temperature is varied by changing the number $N_t$. The latter approach avoids possible misinterpretations of the results due to the additive renormalization of the fermion condensate. We have also investigated the chiral transition using the gradient flow as an independent way to deal with renormalization.

Within the accuracy of the measurement our results indicate a coincidence of the chiral and deconfinement transitions. This is quite remarkable and might hint towards a more general relation between the two transition points. This observation might appear to be in contradiction with previous investigations of theories with fermions in the adjoint representation, but these have been done with a larger number of fermions. Due to the vicinity of the conformal window, investigations of these theories suffer from potential bulk transitions and large finite size effects.

Interestingly, our observation is in agreement with conjectures based on string theory in large $N_c$ limit. Unlike the $\mathcal{N}=4$ case, $\mathcal{N}=1$ SYM is not immediately related to a gauge/gravity duality, but it is possible to incorporate the breaking of supersymmetries in certain brane configurations \cite{Witten:1997ep}. This has lead to the conjecture that chiral and deconfinement transitions should coincide. A nice picture for the description of the $N_c=3$ case in an effective theory has been found in \cite{Campos:1998db}. In this reference the formation of high and low temperature domain walls was related to effects in solid state physics.

\subsection{Compactified theory}
The ground state structure of $\mathcal{N}=1$ SYM has already been investigated in the early seminal paper \cite{Witten:1982df}. In this work, the different vacua of the theory at strong coupling have been identified by means of a weak-coupling and semiclassical analysis. The fact that strongly coupled ground state physics is accessible with analytic approaches is a remarkable feature of supersymmetric theories. It is derived from the invariance of a twisted partition function, the Witten index. In a path integral formulation the twisted partition function corresponds to a change of fermion boundary conditions. The partition function of the canonical ensemble has antiperiodic boundary conditions for fermions and periodic boundary conditions for bosons in the time direction, which implies a breaking of supersymmetry at finite temperatures. The twisted partition function has periodic fermion boundary conditions, which has the physical interpretation of a gauge theory on the compactified manifold $S^1\times R^3$.

Due to a cancellation of fermion and boson energy eigenstates, the twisted partition function projects on ground state information even at small compactifications. This property is sometimes called ``distillation'', with possible extensions even beyond supersymmetric theories \cite{Dunne:2018hog}. It implies the invariance of the twisted partition function even at small compactifications, where semiclassical approximation applies. In fact $\mathcal{N}=1$ SYM can be seen as the cleanest realization of a continuity between the strong coupling and the semiclassical regime. This continuity has been the aim of many alternative constructions like deformations of Yang-Mills theories by Polyakov line operators etc. The ultimate goal of these investigations is to establish connections to ground state physics of Yang-Mills theory and QCD \cite{Shifman:2007ce,Unsal:2007vu,Misumi:2014raa,Aitken:2017ayq}. Numerical lattice simulations are required to establish and crosscheck these relations.

We have investigated $\mathcal{N}=1$ SYM on $S^1\times R^3$ and verified the absence of a deconfinement transition at small enough gluino masses \cite{Bergner:2014dua}. The absence of phase transitions can also be understood as an additional verification of the effective restoration of supersymmetry on the lattice. The temperature derivative of the partition function, i.~e.\ the twisted version of an energy density, measures a weighted difference between fermion and boson energy eigenstates. It is found to be zero for the compactification range at our simulation parameters \cite{Bergner:2015cqa}. 

The fermion discretization has again an important impact on the correct realization of the compactified regime. The theory at small compactification radius with Wilson fermions behaves similar to a theory with additional fermion flavors especially at larger fermion masses in lattice units \cite{Bergner:2018unx}.

Compactified theories with adjoint fermions are also considered in the context of the Hosotani mechanism, and extensions towards the non-perturbative domain have been the motivation for numerical investigations of these theories in \cite{Cossu:2013ora}. In these studies, there was a larger number of fermions than in SYM and a staggered fermion discretization was chosen, making it hard to compare with our results. A confined regime at a small compactification radius was observed. In contrast to the SYM, this confined regime at a small radius is never connected to the large radius confined regime in the investigated parameter range. There seems to always be a deconfined intermediate phase between the two confined phases instead of a continuity.

\section{Towards the conformal window with fermions in the adjoint representation}
$\mathcal{N}=1$ SYM corresponds to Yang-Mills theory coupled to the smallest possible fermion content in the adjoint representation. There has been a significant effort to investigate theories with a larger number of adjoint flavors as well. One original motivation was the walking technicolor scenario. In this scenario the Higgs sector of the Standard Model is replaced by an additional strongly coupled theory. In order to be consistent with electroweak precision data, the theory is required to have a slow running of the gauge coupling with a large mass anomalous dimension over a wide range of energies. It has been conjectured that a theory in the near conformal or walking regime, explained below, could be a realization of such a scenario. In the fundamental representation, a large number of fermions would be needed to approach such a regime, which implies also a large number of additional states that need to be found in experimental data. The so-called S-parameter \cite{Peskin:1990zt}, which is required to be small due to experimental bounds, also scales with the number of fermions. It is therefore a natural step to consider instead theories with adjoint fermions since in this case the perturbative estimates predict near conformality with a much smaller fermion content. Further details about phenomenological applications can be found in review articles like \cite{Hill:2002ap,Cacciapaglia:2020kgq}.

\begin{figure}[t]
\begin{center}  
\includegraphics[width=0.45\textwidth]{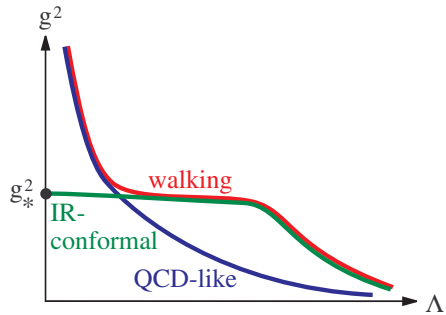}
\end{center}
\caption{\label{muenster-walking}
A sketch of the behaviour of the coupling strength $g^2$ as a function of
the energy scale $\Lambda$ for QCD-like, walking, and infrared conformal
theories.}
\end{figure}

Yang-Mills theory without or with only a small number of fermions has the well known QCD-like running of the gauge coupling from asymptotic freedom at high energies to confinement at low energies. The derivative of the dependence of the strong coupling on the scale is given by the beta function. The perturbative expansion of the beta function for SU($N_c$) gauge theory with $N_f$ fermions in representation $R$ is \cite{Caswell:1974gg,Dietrich:2006cm} 
\begin{align}
\beta_s(g^2)&=-b_0\frac{g^3}{16\pi^2}-b_1\frac{g^5}{(16\pi^2)^2}+\mathcal{O}(g^7)\label{eq:beta}\\
b_0&=\frac{11}{3}N_c-\frac43 N_f \frac{C_2(R) d(R)}{N_c^2-1}\nonumber\\
b_1&=\frac{34}{3}N_c^2-\left(\frac{20}{3}N_c \frac{C_2(R) d(R)}{N_c^2-1}+4 \frac{C_2^2(R) d(R)}{N_c^2-1}\right)N_f\; ,\nonumber
\end{align}
where $C_2(R)$ is the quadratic Casimir operator and $d(R)$ the dimension of representation $R$. For the fundamental representation the constants are $C_2(R)=\frac{N_c^2-1}{2N_c}$, $d(R)=N_c$, and for the adjoint one $C_2(R)=N_c$, $d(R)=N_c^2-1$. 
In this perturbative beta function the fermion and gauge field contributions enter with the opposite sign. In QCD both $b_1$ and $b_0$ are positive and the coupling increases at lower energy scales. With a large number of fermions, the beta function has a positive sign even at small couplings and asymptotic freedom is lost. In an intermediate range for the number of fermions the two coefficients come with opposite signs, which leads to a sign change of the beta function at a certain value of the gauge coupling. This implies the existence of an infrared (IR) fixed point at which the running of the gauge coupling vanishes \cite{Banks:1981nn}. Theories with an IR fixed point show scale-invariant (conformal) behaviour at large distances, while walking theories have a nearby IR fixed point. In both cases the coupling strength varies only slowly over a large range of scales, in contrast to the running coupling of QCD, see Figure \ref{muenster-walking}.

Theories with sufficient fermion content to show an IR fixed point are said to be inside the conformal window. Theories in the conformal window show a behavior much different from standard QCD. Close to the fixed point, at sufficiently small fermion masses $m_0$, the mass of any particle $M$ should scale to zero as $M\propto m_0^{1/\gamma}$ with the same dependence given by the mass anomalous dimension $\gamma$. At vanishing fermion masses, there is no chiral symmetry breaking and no mass gap.
From a general theoretical point of view and also for phenomenological considerations, it is very interesting to investigate such completely different regimes in the landscape of gauge theories. 

So far we have discussed only the perturbative estimates, which are insufficient to show whether such a scenario can be realized.
At the lower end of the conformal window (with respect to the number of fermions), the fixed point is at strong couplings and numerical simulations are required to investigate the infrared properties of the theory. 

Equation \eqref{eq:beta} shows that for fermions in the adjoint representation, the conformal window starts at a much lower number of fermion fields compared to fermions in the fundamental representation. As explained above, this makes it easier to fulfill experimental constraints and has been one of the main motivations for an investigation of Yang-Mills theories coupled to adjoint fermions.

\subsection{Two Dirac flavors: Minimal walking technicolor}
\begin{figure}
	\begin{center}
	\begin{subfigure}[b]{0.49\textwidth}
		\centering
		\includegraphics[width=\textwidth]{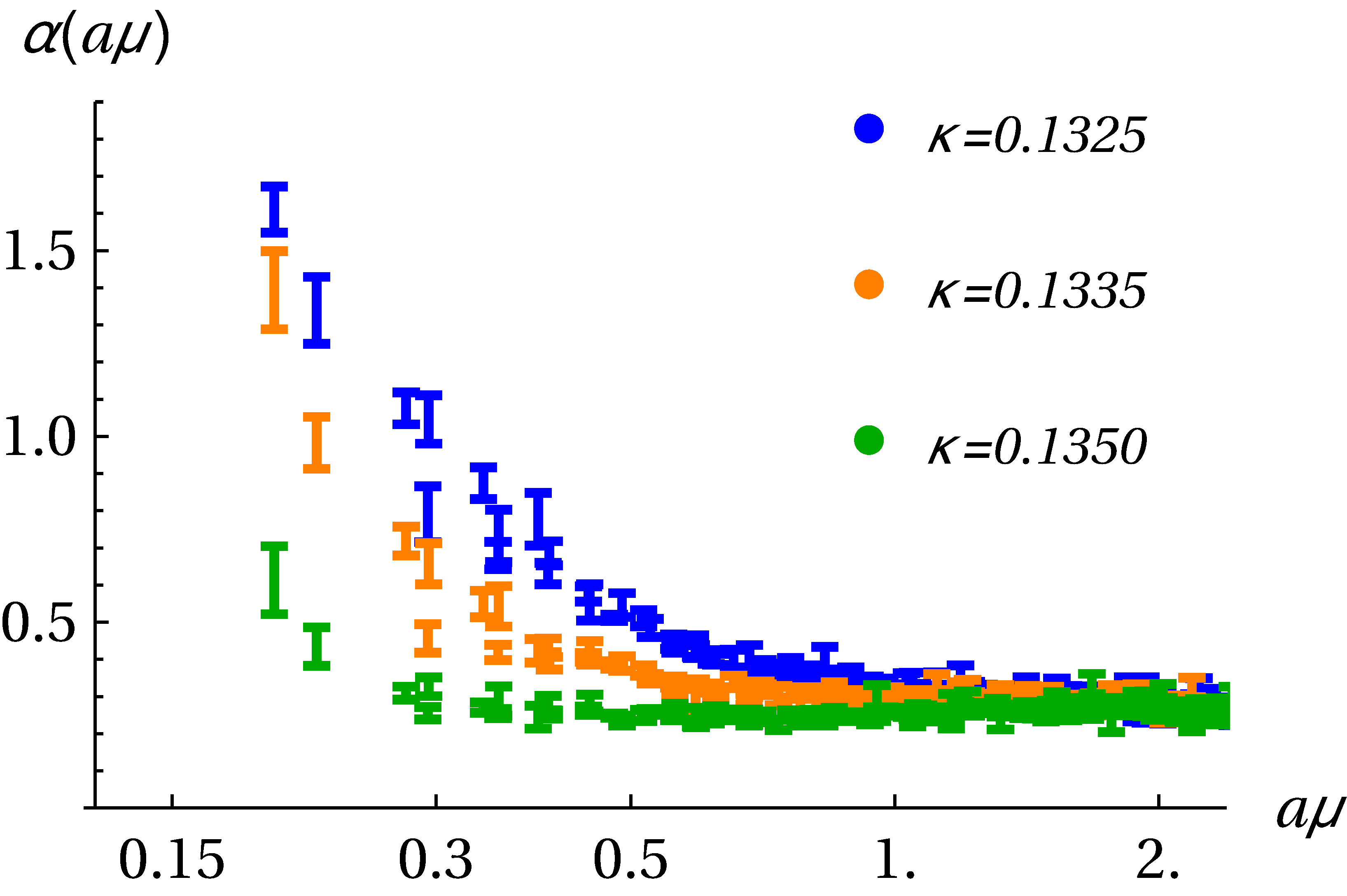}
		\caption{Running of the coupling}
		\label{fig:MWTRunningBeta}
	\end{subfigure}
	\hfill
	\begin{subfigure}[b]{0.49\textwidth}
		\centering
		\includegraphics[width=\textwidth]{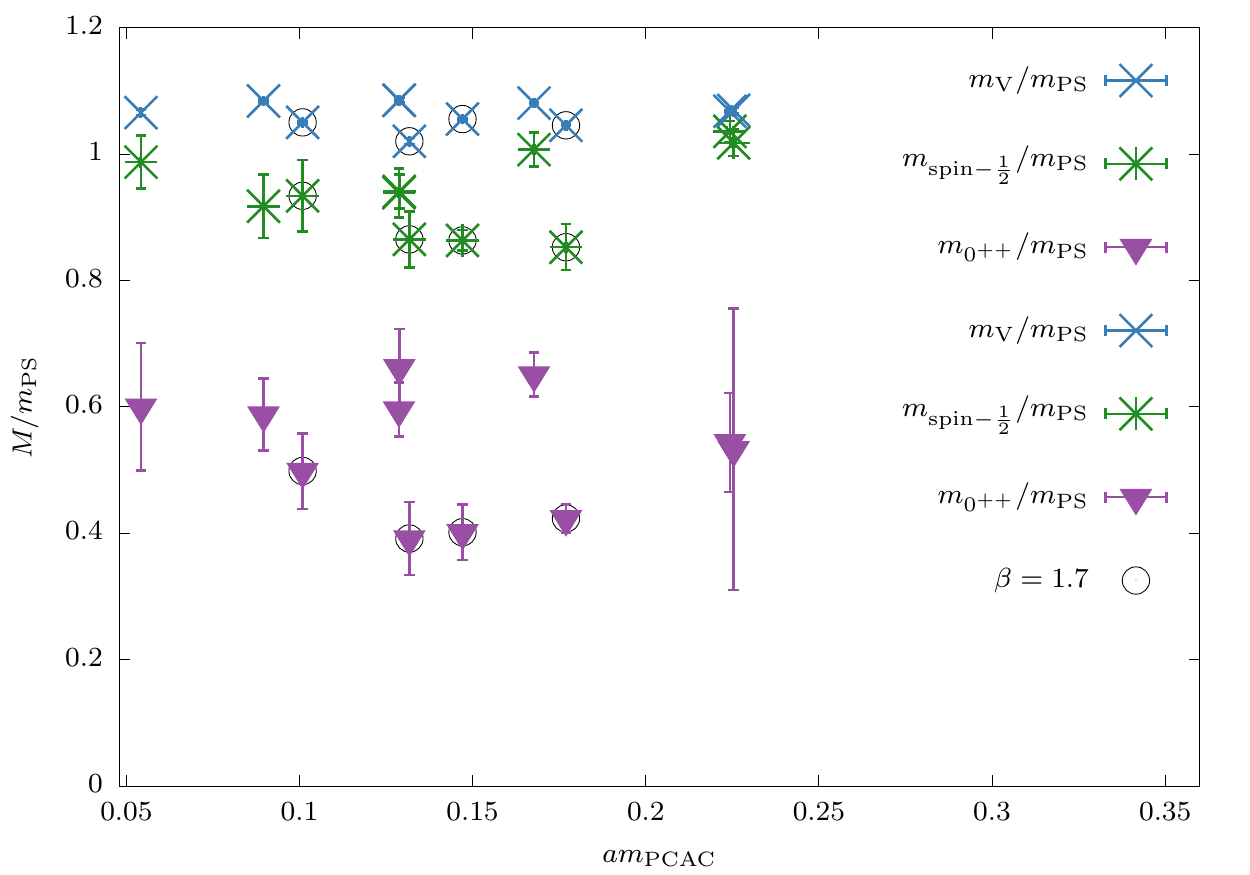}
		\caption{Bound state spectrum}
		\label{fig:MWTParticle}
	\end{subfigure}
	\end{center}
	\caption{Numerical results for properties of adjoint QCD with two Dirac flavors.\\Left hand side: Running of the coupling for increasing $\kappa$ corresponding to a decreasing fermion mass, determined from gluon and ghost propagators, see \cite{Bergner:2017ytp} for further details.\\Right hand side: masses of bound state particles in units of the pseudoscalar meson mass (pNGb like pion in QCD) as a function of the partially conserved axial current mass (proportional to renormalized fermion mass). Two different values of the gauge coupling are combined. $m_V$ is the vector meson mass, $0^{++}$ indicates the scalar glueball mass, $m_{\text{spin}-\frac12}$ corresponds to the gluino-glue mass in SYM. The data is obtained at a inverse gauge coupling of $\beta=1.5$ with some additional points at $\beta=1.7$, see \cite{Bergner:2016hip} for further details.  }
	\label{fig:MWT}
\end{figure}

The most well studied theory in this context is SU(2) adjoint QCD with two Dirac flavors (Minimal Walking Technicolor, MWT). In this case, the flavor content is just large enough to enable a coupling to the Standard Model, and first perturbative estimates have concluded that it should be near-conformal. A rather large number of numerical investigations have been carried out for this theory, including the particle spectrum, the running of the gauge coupling, and the mass anomalous dimension \cite{Catterall:2007yx,Catterall:2008qk,Hietanen:2009zz,DeGrand:2011qd,Patella:2012da,Rantaharju:2015yva,Bergner:2016hip}. These studies have indeed found good agreement with infrared conformality. As expected, the masses scale to zero in a way determined by a common mass anomalous dimension, which implies constant mass ratios. The lightest particle is a scalar state and not the pseudo Nambu--Goldstone bosons (pNGbs). An example of the results is shown in Figure \ref{fig:MWT}. The anomalous dimension obtained in \cite{Patella:2012da} seems to be quite low compared to the expectations for a walking technicolor scenario.

Despite all of the numerical investigations, there are very relevant unresolved puzzles even in case of this theory. It is indeed quite remarkable that the scalar state is so much lighter than the pNGbs. It might be that it is a realization of a dilaton generated by the breaking of conformal invariance \cite{Yamawaki:1985zg}. It seems that proposed descriptions of near-conformal theories in terms of effective actions, which try to combine chiral perturbation theory with a light scalar state \cite{Golterman:2016lsd}, are, however, not really applicable. Instead, it seems that the mass ratios are fixed at values corresponding to a theory with heavy fermion masses. In that sense, the theory always looks like being in a heavy mass regime. In this case, alternative effective descriptions have to be found. Alternatively, it might imply that the fermion masses have so far not been sufficiently low to investigate this theory.

A second puzzle is the remaining dependence of the obtained values for the fixed-point mass anomalous dimension $\gamma$ on the gauge coupling. In theory, one would expect a universal value for an IR conformal theory, but, in practice, values in the range of $\gamma=0.2$ to $0.5$ have been observed and the value changes for different values of the inverse gauge coupling $\beta$ \cite{Bergner:2016hip}. Possible interpretations are lattice artifacts at smaller $\beta$, or one might be in a different mass regime at larger $\beta$. Large volume studies and considerations of different discretizations could help to resolve the tensions within current data.

While most investigations are performed with standard methods, some interesting alternative approaches have also been applied. An example are investigations in the large $N_c$ limit. From the perturbative analysis, it is expected that the boundary of the conformal window and the mass anomalous dimension show only a small $N_c$ dependence. Therefore, methods derived in the large $N_c$ limit can provide insights about the general conformal behavior. Such kinds of investigations were presented in \cite{Catterall:2010gx,Bringoltz:2011by,GarciaPerez:2015rda} and also indicate a conformal scenario for the theory.

\subsection{Theories between two Dirac flavors and $\mathcal{N}=1$ supersymmetric Yang-Mills theory}
In order to investigate more of the general landscape of theories and help to understand better the case of two adjoint Dirac flavors, also SU(2) with one Dirac flavor has been considered \cite{Athenodorou:2014eua}. Later on the picture has been completed by a study with three Majorana flavors \cite{Bergner:2017gzw}. The main aim of the first studies has been the determination of the lower boundary of the conformal window and whether large values of $\gamma$ required by the walking technicolor scenario can be achieved in any kind of strongly interacting theory. Indeed, while SYM shows confining properties, the theories with larger number of adjoint fermions show properties consistent with an IR fixed point. The obtained estimates for the mass anomalous dimension rise with decreasing fermion content starting from the two adjoint Dirac flavor theory.

SU(2) Yang-Mills theory with one adjoint Dirac flavor has additional interesting applications and relations to other theories. Recently the theory has gained interest due to the different possible conjectured infrared scenarios. It could be a confining theory with chiral symmetry breaking, it could be conformal or near conformal, but it could also follow a completely different infrared scenario, conjectured based on 't Hooft anomaly matching conditions \cite{Anber:2018iof}. It corresponds to $\mathcal{N}=2$ supersymmetric Yang-Mills theory without scalar fields and can hence be understood from that perspective as a limiting case with broken supersymmetry. 
This has been the motivations for more recent studies on the lattice \cite{Bi:2019gle,Athenodorou:2021wom,Bergner:2022hoo}, but so far the infrared scenario has not been completely resolved. The theory has a light scalar, and mass ratios remain almost constant, whereas a chiral fit does not provide a good description of the data. Nevertheless there is an unresolved dependence of the anomalous dimension on the gauge coupling, and the data is not completely consistent with scale invariance. A first investigation with overlap fermions shows instead indications for chiral symmetry breaking and confinement \cite{Bergner:2022hoo}.  

From a technical point of few, theories with an even number of Majorana flavors are easier to simulate since they don't require the evaluation of the Pfaffian. The case of four Majorana flavors, corresponding to two adjoint Dirac fermions, is even simpler since an ordinary HMC algorithm can be applied instead of the RHMC. Nevertheless we have also been able to investigate SU(2) Yang-Mills theory with three Majorana flavors \cite{Bergner:2017gzw} based on our experience with SYM. The obtained data show constant mass ratios and a scaling of the distribution of lowest eigenvalues of the Dirac operator, which is consistent with scale invariance. 

\subsection{Combining adjoint and fundamental fermion fields}
The landscape of considered theories can be further extended if both, fundamental and adjoint fermion fields are coupled to the gauge theory. The first motivation to consider such kind of theories arose again in the context of walking technicolor theories. The primary goal was to minimize the S-parameter to avoid tensions with the experimental bounds. The theory obtained in this consideration consists of SU(2) gauge theory coupled to one Dirac fermion in the adjoint and two Dirac fermions in the fundamental representation \cite{Ryttov:2008xe}.

Considering the above numerical results, which indicate a conformal scenario with a rather small mass anomalous dimension for SU(2) with two adjoint Dirac flavors, it seems natural to extend the one adjoint Dirac flavor case such that it can be coupled to the Standard Model. The most natural way to achieve this is indeed an extension by two fundamental Dirac flavors. This can also be considered as a modification of the SU(2) gauge theory with two Dirac flavors considered as a candidate for composite Higgs theories. There is indeed not a clear distinction in such a theory between the composite Higgs scenario, where the electroweak symmetry breaking appears due to a coupling to a scalar pNGb, and a technicolor theory, where the fermion condensate induces the breaking and a light scalar is induced by a different mechanism \cite{Cacciapaglia:2014uja}. A recent review of SU(2) Yang-Mills theory with two Dirac fermions and possible extensions including some numerical lattice results can be found in \cite{Cacciapaglia:2020kgq}.

The mixed adjoint and fundamental fermion fields are also the first step towards investigations of $\mathcal{N}=1$ supersymmetric QCD (SQCD). This theory consists of a gauge sector represented by SYM coupled to matter supermultiplets with fermionic quark and bosonic squark fields in the fundamental representation. The limiting case, in which the scalars become heavy is a theory with adjoint and fundamental fermions. In other considerations, the adjoint fermions are coupled to QCD like theories with fundamental quarks to investigate their non-perturbative properties and find relations to the semiclassical regime \cite{Kanazawa:2019tnf}. 

Recently theories with two different fermion representations have been considered in non-perturbative lattice simulations \cite{Ayyar:2017qdf,Cossu:2019hse}. The low energy effective theory description for this case has also been derived recently \cite{DeGrand:2016pgq}. The case of SU(2) gauge theory with fundamental and adjoint fermions has been studied for the first time in \cite{Bergner:2020mwl}. This first study basically considered only the question whether a chiral confining or a conformal scenario is observed for SU(2) Yang-Mills theory with two fundamental and one adjoint Dirac flavors. Based on the limited data, the results rather indicate that the theory is confining.

\section{Towards supersymmetric QCD and extended supersymmetry}
The field content of the theories explored in previous sections is already the same as that of supersymmetric QCD, $\mathcal{N}=2$, and even $\mathcal{N}=4$ SYM if scalar fields are neglected. Scalar fields can be incorporated as a minor modification of the theory as long as they remain in a heavy mass regime. Once the masses are lowered, and the scalar potential as well as Yukawa interactions become relevant, a highly non-trivial phase diagram as a function of the scalar couplings appears, with additional Higgs phases and different vacuum solutions due to flat directions of the scalar potential. This can be explained considering, for example, the additional terms required to couple the matter multiplet in supersymmetric QCD
\begin{align}
\mathcal{L}_{SQCD}&=\mathcal{L}_{SYM}+\mathcal{L}_{kin}+ i\sqrt{2}g \bar{\lambda}^a\left(\Phi_1^\dag   T^a P_{+}+\Phi_2 T^aP_- \right) \psi\nonumber\\
& -i\sqrt{2}g \bar{\psi}\left(P_- T^a \Phi_1 + P_+T^a\Phi_2^\dag \right) \lambda^a
+\frac{g^2}{2} \left( \Phi_1^\dag T^a \Phi_1-\Phi_2^\dag T^a \Phi_2\right)^2,
\label{eq:SQCD}
\end{align}
where $\mathcal{L}_{SYM}$ is the gauge part represented by $\mathcal{N}=1$ SYM and $\mathcal{L}_{kin}$ is the usual kinetic term for the scalar fields $\Phi_1$, $\Phi_2$, and the fermion $\psi$. These complex scalar fields are like the fermion $\psi$ in the fundamental representation of the gauge group. $T^a$ are the generators of the gauge group and $\lambda$ denotes the gluino field. Like in many other supersymmetric theories, the scalar potential defined by the last term in \eqref{eq:SQCD} has flat valleys in the scalar potential, which means a continuous space of minima. In simpler theories without supersymmetry, such flat directions are usually lifted by scalar quantum fluctuations, but in supersymmetric gauge theories these are often canceled by fermionic contributions. In the example of \eqref{eq:SQCD}, which means one-flavor SU($N_c$) SQCD in the massless limit, the vacuum is even expected to become unstable once quantum corrections are included. This means that the scalar field is driven towards infinity along the flat directions, see for example \cite{Terning:2006bq} for an introduction.

The scalar fields lead to considerable complications once numerical lattice simulations are considered. One important advantage of supersymmetry is the fact that it induces severe restrictions on the scalar part of the action. This is basically the main reason why certain small parameters appear in a more natural way in supersymmetric versions of theories with scalar fields. This symmetry is, however, explicitly broken on the lattice. Therefore one has to tune the parameters in a larger space of scalar couplings compared to the ones consistent with supersymmetry. This seems to be difficult, but more precise considerations are required in order to estimate the feasibility. Some work has already been done based on lattice perturbation theory and exploratory numerical simulations \cite{Elliott:2008jp,Giedt:2009yd,Costa:2017rht,Wellegehausen:2018opt,Bergner:2018znw}. Note that in $\mathcal{N}=4$ SYM one can reduce the tuning problem by considering a twisted version of the theory, which allows to keep part of the supersymmetry preserved on the lattice, see \cite{Catterall:2009it} for a review.

Besides these basic difficulties, there are also practical challenges. One example are sign problems. After integrating out the fermions, the Pfaffian of the Dirac operator includes Yukawa interactions with scalar fields. Due to fluctuations of scalar fields, bounds on the Pfaffian can only be established in limiting cases. With sign fluctuations the path integral measure can not be interpreted as a statistical probability and standard Monte-Carlo methods fail. The sign can be included by reweighting at the expense of larger uncertainties, which might even render the method unreliable.

\section{Adjoint dark matter}
Extensions of the Standard Model with an additional strongly interacting sector coupled to the electroweak/Higgs sector have been considered in order to find more natural explanations of the Higgs mass. This has been one the motivations for supersymmetric and composite Higgs theories. Such extensions of the Standard Model can, however, be considered in a more general context. For example, another application is to find a strongly interacting theory that describes the dark matter found in astronomical observations. In this context, a rather broad scan of possible theories is required to find realistic candidates and experimental signatures; see \cite{Kribs:2016cew} for a review in the context of numerical lattice simulations. 

In most studies, QCD like gauge theories with fermions in the fundamental representation are considered.  However, adjoint fermions coupled to gauge fields provide an interesting alternative scenario with unique features, which can be checked in experimental observations. The additional feature considered in phenomenological considerations is the appearance of composite bound states of an adjoint fermion with gluons, sometimes called gluequark, which in SYM are the gluino-glue states. A comparison to other strongly interacting dark matter models can be found in \cite{Cline:2021itd}. This kind of dark matter has been considered as glueballinos in the context of a hidden supersymmetric sector and explanation of astronomical X-ray signatures \cite{Boddy:2014qxa}. A systematic study in \cite{Contino:2018crt} shows that the fermions can be even very heavy in such a dark matter scenario. 
These theories also have quite unique signatures concerning phase transitions in the early universe and possible gravitational wave signals \cite{Reichert:2021cvs}. It would be interesting to complement these studies with numerical lattice simulations.

\section{Conclusions}
We have summarized investigations and applications of gauge theories coupled to fermions in the adjoint representation. The adjoint matter provides a non-trivial extension of the landscape describing realizations of strong interactions. This extension includes supersymmetric gauge theories. 

The minimal matter content of these theories is realized by $\mathcal{N}=1$ SYM. We have reviewed that this theory shows interesting properties due to supersymmetry and a rich phase structure with $N_c$ vacua related to the gluino condensate and the deconfinement transition at finite temperature. The fact that the non-perturbative properties of this theory have become accessible by current lattice simulations could encourage additional considerations and investigations.

Regarding the infrared limit, $\mathcal{N}=1$ SYM is similar to QCD with confinement and asymptotic freedom. Increasing the number of adjoint fermions, the running of the gauge coupling changes considerably. For SU(2) adjoint QCD with two Dirac flavors, all numerical simulations show a consistent indication of an infrared conformal fixed point, which is quite remarkable compared to other theories studied in this context. However, some puzzles remain and in particular the influence of lattice artefacts requires further studies.
It is still difficult to determine the precise lower bound of the conformal window. This makes it hard to arrive at conclusive results for the interesting case of one-flavor adjoint QCD.

We have extended the landscape of considered theories to include also a combination of adjoint and fundamental matter as well as scalar fields. Supersymmetric QCD and theories with extended supersymmetry are part of this larger space of theories. This space of Higgs-Yukawa models with different fermion content combines the difficulties observed in numerical studies of Higgs models regarding different vacuum solutions and phase transitions with the challenges due to the fermion content. For this reason numerical investigation of these theories is still at its infancy. Nevertheless this is still the only way to get more general insights for the non-perturbative sector of these theories.

We briefly mentioned further applications of gauge theories with adjoint matter besides supersymmetric and composite Higgs extensions of the Standard Model. The theories provide interesting alternative dark matter scenarios. Moreover, they are the foundation of theoretical investigations and new approaches for an understanding of strong interactions.

The considered theories all come with their own technical challenges. The realization of supersymmetry on the lattice requires specific considerations, and the bound states of SYM are also difficult to measure. Larger number of fermions lead to a near-conformal theory, which has different finite-size effects and possible bulk phases. Nevertheless, over the past few decades, enough experience has been obtained with the simulations of these theories to allow for reliable numerical investigations.
The space of theories with scalar fields has only been partially explored, and further work is required, especially regarding the supersymmetric limit of these theories.

\section*{Acknowledgements}\vspace*{-0.1cm}
We thank D.~Schaich for discussions. G.~B.\ is funded by the Deutsche Forschungsgemeinschaft (DFG) under Grant No.~432299911 and 431842497.
We gratefully acknowledge the Gauss Centre for Supercomputing e.~V.\ (www.gauss-centre.eu) for funding this project by providing computing time on the GCS Supercomputer SuperMUC-NG at Leibniz Supercomputing Centre (www.lrz.de). Further computer time has been provided by the PALMA HPC cluster of the University of Münster.
\newpage
\begin{adjustwidth}{-\extralength}{0cm}
	\reftitle{References}
	\bibliography{paper}
\end{adjustwidth}
\end{document}